# Suppression of transport spin-polarization of surface states with emergence of ferromagnetism in Mn-doped $Bi_2Se_3$

Suman Kamboj[1], Shekhar Das[1], Anshu Sirohi[1], Rajeswari Roy Chowdhury[1],*
Sirshendu Gayen[1], Vishal K. Maurya[2], Satyabrata Patnaik[2], and Goutam Sheet[1]†
[1]*Department of Physical Sciences, Indian Institute of Science Education and Research Mohali, Sector 81, S. A. S. Nagar, Manauli, PO: 140306, India and*
[2]*School of Physical Sciences, Jawaharlal Nehru University, New Delhi, PO: 110067, India*

The surface states of topological insulators (TI) are protected by time reversal symmetry and they display intrinsic spin helicity where the momentum of the charge carriers decides their spin states. As a consequence, a current injected through the surface states becomes spin polarized and this transport spin-polarization leads to a proportionate suppression of Andreev reflection in superconductor/TI junctions. Here we show that upon doping $Bi_2Se_3$ with Mn, the transport spin-polarization is seen to be monotonically suppressed. The parent compound $Bi_2Se_3$ is found to exhibit a transport spin-polarization of about 63% whereas crystals with 10% Mn doping show transport spin-polarization of about 48%. This suppression is accompanied by an increasing ferromagnetic order of the crystals with Mn doping. Scanning tunneling spectroscopy shows that the topological protection of the surface states reduce due to Mn doping. The net measured transport spin-polarization is due to a competition of this effect with the increased magnetization on Mn doping. The present results provide important insights for the choice of magnetic topological insulators for spintronic applications.

Strong spin-orbit coupling leads to bulk band inversion in topological insulators (TI) and as a consequence, non-trivial conducting states comprising of massless Dirac fermions emerge at the surfaces of TIs [1–5]. Such conducting states at the surface of a gapped bulk are protected by time reversal symmetry. The Dirac charge particles corresponding to these surface states have spin locked with its momentum. This so-called spin-momentum locking leads to large transport spin-polarization of the surface states [6–9]. If the time reversal symmetry can be broken either by applying an external magnetic field or by doping a TI with magnetic dopants in a controlled fashion, it may be possible to control the transport spin-polarization of the surface states. From theoretical calculations it is also known that magnetic dopants can lead to opening of a gap at the Dirac point in 3D TIs and with such doping, the TIs may exhibit ferromagnetic order with perpendicular magnetic anisotropy facilitated by Ruderman-Kittel-Kasuya-Yosida (RKKY) exchange [10–16]. Ferromagnetism in TIs may lead to wider variety of exotic physical phenomena, such as the anomalous quantum Hall effect and magneto-electric effects [17, 18]. In this paper, we present scanning tunneling spectroscopy and spin polarized Andreev Reflection spectroscopy and demonstrate that with doping of magnetic atoms (Mn), as a ferromagnetic phase emerges, the topological protection of the system breaks down and the effective transport spin polarization at the surface decreases. This decrease in spin polarization with magnetic doping can be attributed to a competition between the spin polarization originating from the spin momentum locked surface states, which is lowered in the doped system and the emerging magnetization of the system that increases with doping.

High quality single crystals of $Mn_xBi_{2-x}Se_3$ ($x$= 0, 0.03, 0.05, 0.1) were cleaved by an *in-situ* cleaver at 80 K under ultra-high vacuum ($10^{-11}$mbar). In order to confirm the quality of the undoped and the doped crystals, we first carried out Low Energy Electron Diffraction (LEED) *in-situ*. As shown in the supplementary materials long range ordering of the atoms in the crystals of $Mn_xBi_{2-x}Se_3$ remain unchanged with $x$. All the crystals show hexagonal LEED pattern. No extra features due to any other impurities or clusters could be found in the LEED patterns. The absence of any additional impurity phase was also confirmed by XRD analysis. The crystals were then transferred to the scanning tunneling microscope (STM) measurement head kept at low temperature (Unisoku system with RHK R9 controller, working down to 300 mK).

Figure 1 (a) shows atomic resolution image of the $Bi_2Se_3$ surface captured at 17 K where clover shaped/triangular defects are visible. The defects are randomly distributed throughout the crystal surface. The triangular defects (Figure 1 (a)) are known to be associated with Se-vacancy in the binary selenide family of materials [19, 20]. The lower end inset of Figure 1(a) shows enlarged view of one such triangular defect state. The top right inset of Figure 1 (a) shows zoomed in view of a small defect free area in the undoped $Bi_2Se_3$ sample taken at 17 K exhibiting periodic arrangement of atoms. After confirming the pristine nature of the surface through atomic resolution imaging, we performed local scanning tunneling spectroscopy (STS) at several points on the surface of the crystals using the STM tip. A representative STS spectrum on $Bi_2Se_3$ is shown in Figure 1(b) where a “Dirac cone” is clearly observed with the Dirac point at 100 meV below the Fermi energy. Usually in topological insulators the concentration of defects play an important role in determining the position of the Dirac point energy in a crystal with respect to its Fermi energy. As shown in Figure 1 (c), Mn doping in $Bi_2Se_3$ resulted in the disappearance of the “V” shape of the spectrum and resulted in a gap like structure with the top of the valence band at -400 meV and bottom of the conduction band at 200 meV. This may also indicate gradual suppression of topological protection due to Mn doping. This gap has been observed to exist over a larger span of energy with increasing Mn doping as can be clearly seen from Figure 1 (d). The enhanced contribution of the bulk with Mn doping may also be responsible for the gap like structure.

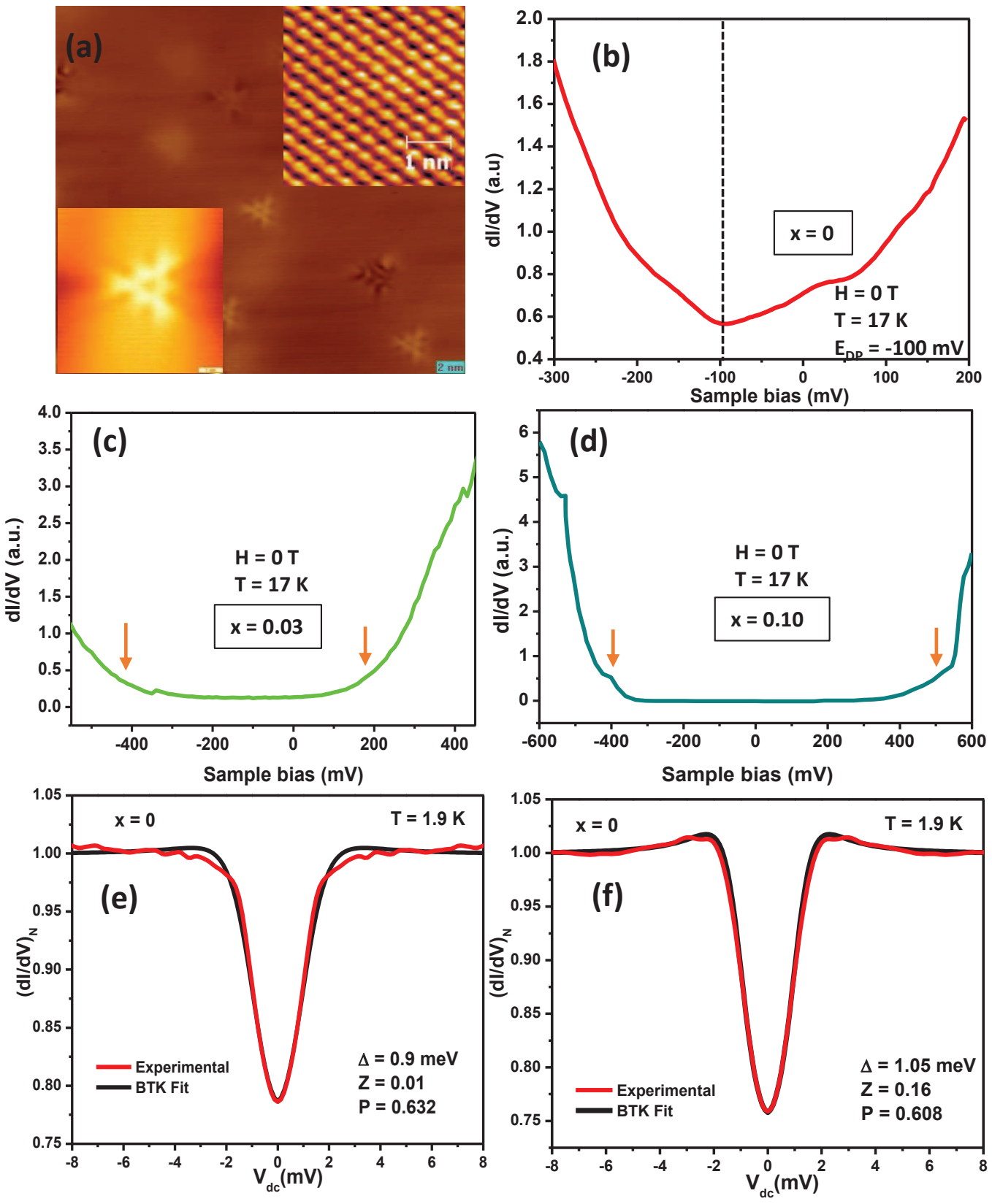

Figure 1: (a) Atomic resolution image of the cleaved surface of $Bi_2Se_3$. Inset shows an enlarged view of one defect state. (b) A differential conductance spectrum measured by STS at 17 K. The Dirac point is at -100 mV. STS conductance spectrum for the $Mn_xBi_{2-x}Se_3$ sample with (c) $x = 0.03$ and (d) $x = 0.10$. The arrows indicate the bottom of conduction band (right) and the top of valence band (left), respectively. (e), (f) Normalized dI/dV spectra for point-contacts on $Mn_xBi_{2-x}Se_3$ with $x$=0 using a Nb tip. The black lines show BTK fits with spin-polarization included.

In recent times various studies involving characterization of Mn doped $Bi_{2-x}Se_3$ have been carried out [21–23]. In the past, magnetization measurements on $Mn_xBi_{2-x}Se_3$ exhibited emergence of ferromagnetic correlations with Mn doping [24]. In those studies, though the field dependence of magnetization showed no hysteresis loop for 3 % Mn doping, a hysteresis loop could be clearly observed in the 10 % Mn doped $Bi_2Se_3$ crystals. Further Hall effect studies showed that while the slope of Hall voltage is negative for the undoped and 3 % doped crystals, it changes to positive for higher dopings of Mn [24]. This was attributed to a change in carrier type (from $n$-type to $p$-type).

Based on the above observations, it is important to examine whether the transport spin-polarization can also be controlled in $Bi_2Se_3$ by Mn doping. Point contact Andreev reflection (PCAR) spectroscopy, now a well established technique, was successfully applied to measure the spin-polarization of elemental ferromagnets, like Fe, Co, Ni [25] and ferromagnetic compounds like $SrRuO_3$ [27], CuFeSb [28] *etc.* Recently it was shown PCAR spectroscopy using a superconducting tip on the surface of a topological insulator can also be employed for measuring the transport spin-polarization in TIs [29–31]. This was demonstrated by Borisov *et al.* where they measured spin-polarization of $Bi_2Te_3$ to be 70% which is remarkably higher than the known ferromagnetic metals [29]. However, Andreev reflection (AR) at an interface between a conventional superconductor and a topological material requires closer inspection because of complexities of the coupling between the two phases where the topological phase has a non-trivial spin texture unlike in feromagnets where a well defined order parameter exists. The PCAR spectra on TI's were successfully analyzed by Borisov *et al.* using a modified Blonder-Tinkham-Klapwijk (BTK) model including the role of spin-polarization developed by Strijkers *et al.* [32].

Our PCAR measurements on $Mn_xBi_{2-x}Se_3$ were performed using sharp tips of conventional superconductor niobium (Nb). The measurements involved obtaining the point contact spectra (*i.e.* differential conductance $dI/dV$ *vs.* $V$ curve) for different contacts (with different values of $Z$, $\Delta$, $\Gamma$, $P$) with variation of external parameters like temperature, magnetic field *etc.* The symbols $Z$, $\Delta$, $\Gamma$ and $P$ denote the dimensionless interfacial barrier strength as in the Blonder-Tinkham-Klapwijk (BTK) theory, superconducting gap amplitude, energy scale associated with finite quasi-particle lifetime and percentage transport spin-polarization, respectively [33–36]. Our PCAR spectra were analyzed using the same procedure as in References [6, 29, 32].

In Figure 1(e,f) we show the representative Andreev reflection spectra obtained on undoped $Bi_2Se_3$. It is observed that the zero-bias conductance is low and the peaks (coherence) are shallow indicating strong suppression of AR. It is known that this suppression may originate due to the spin-polarization of the transport current [25, 37]. In a transport experiment like the present one, the relevant quantity is not the absolute spin polarization but the so called "transport spin polarization" which is defined as: P = $(\langle N_\uparrow v_{F\uparrow}\rangle_{FS}$ - $\langle N_\downarrow v_{F\downarrow}\rangle_{FS})/(\langle N_\uparrow v_{F\uparrow}\rangle_{FS}$ + $\langle N_\downarrow v_{F\downarrow}\rangle_{FS})$, where $N_\uparrow$ and $N_\downarrow$ are the density of states (DOS) of the up and down spin channels respectively at the Fermi level and $v_{F\uparrow}$ and $v_{F\downarrow}$ are the respective Fermi velocities. The average is taken over the entire Fermi surface [6, 26]. In order to estimate the degree of spin-polarization all the AR data have been fitted using modified BTK theory following Strijkers *et al.*'s model [32, 33, 38]. The black lines show the fit to the experimentally obtained spectra. Values of different parameters extracted from the fitting of the data are also shown. The superconducting gap Δ ranges between 0.7-1.5 meV for different contacts. The extracted values of P is also seen to slightly depend (linear dependence) on the barrier height Z as can be seen from Figure 5 (a). The solid lines in Figure 5 (a) show linear extrapolation of the Z-dependence of P to Z = 0 which gives the expected intrinsic value of the transport spin-polarization. The intrinsic transport spin-polarization in $Bi_2Se_3$ is obtained to be around 63 % which is consistent with theoretically calculated value reported in the past [39] and is remarkably high compared to some of the ferromagnetic metals like iron (Fe) and Cobalt (Co) both of which possess a Fermi level spin-polarization of about 40 % [25]. The magnitude of spin-polarization obtained in undoped $Bi_2Se_3$ is also comparable to the spin-polarization of other members of the binary chalcogenide family like $Bi_2Te_3$ where a spin polarization of 70 % have been observed [29].

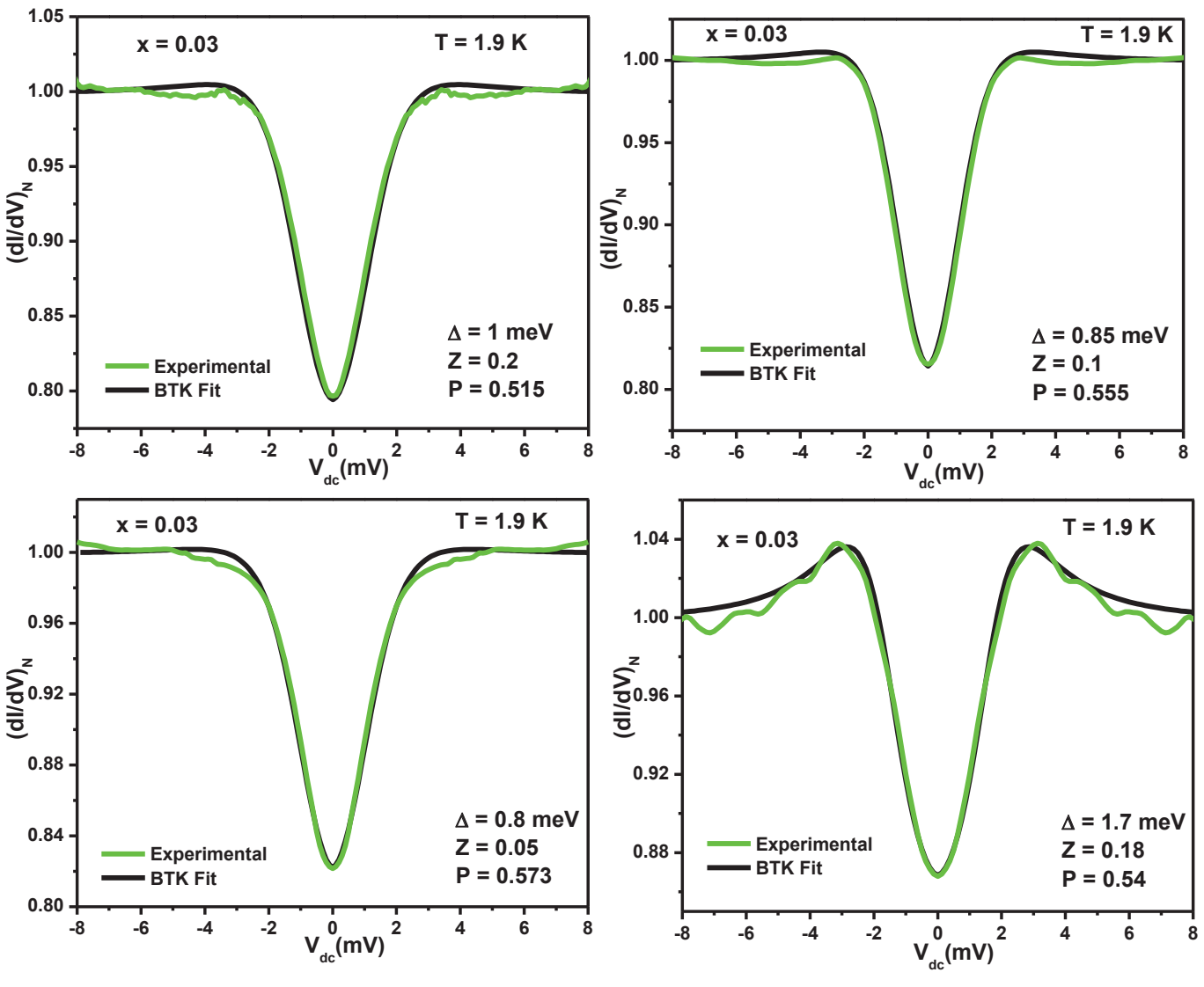

Figure 2: Normalized dI/dV spectra for point-contacts on $Mn_xBi_{2-x}Se_3$ with $x = 0.03$ using a Nb tip. The black lines show BTK fits with spin-polarization included.

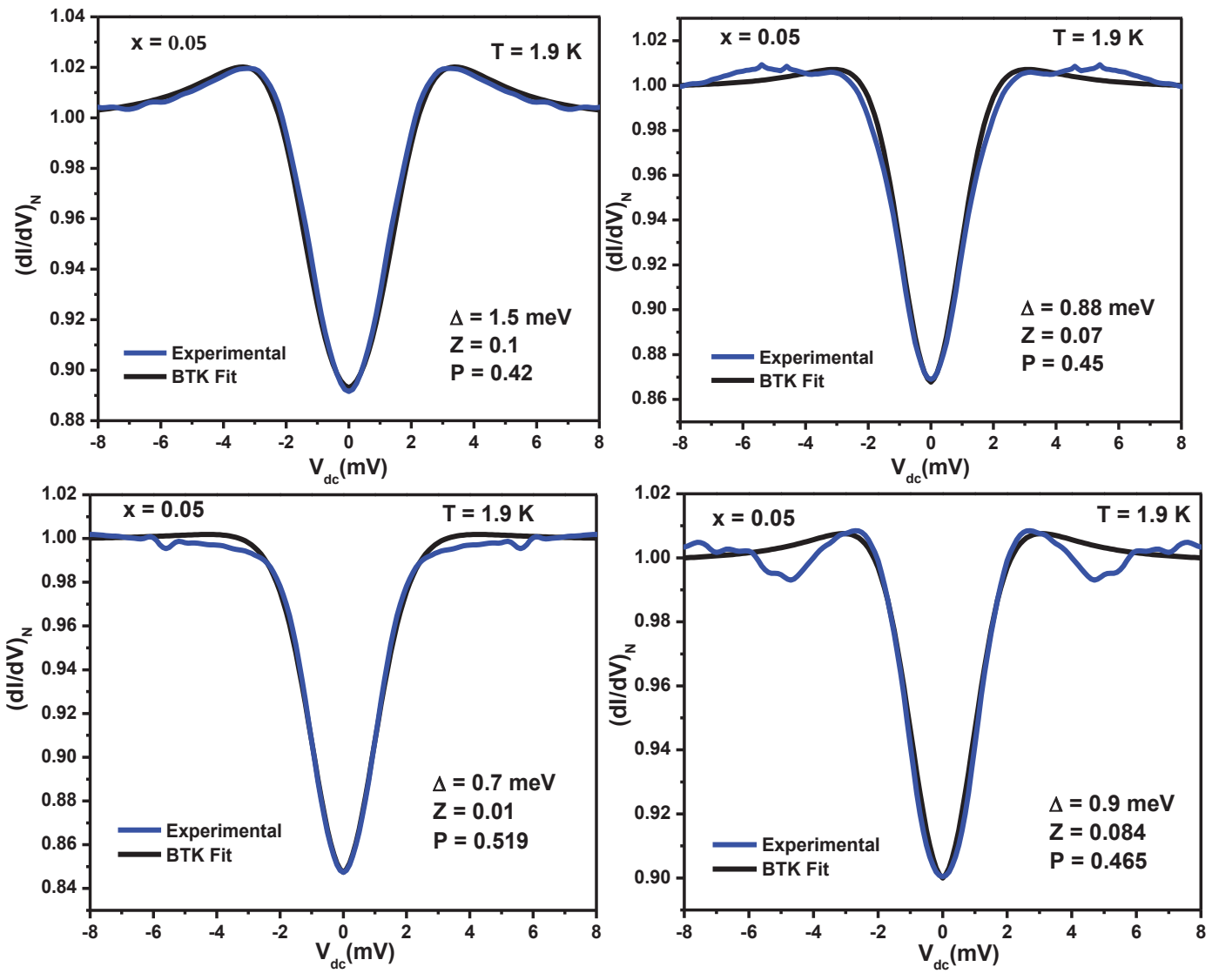

Figure 3: Normalized dI/dV spectra for point-contacts on $Mn_xBi_{2-x}Se_3$ with $x = 0.05$ using a Nb tip. The black lines show BTK fits with spin-polarization included.

The representative PCAR spectra obtained on $Mn_xBi_{2-x}Se_3$ crystals with $x$= 0.03, 0.05 and 0.1 doping are shown in Figure 2, 3 and 4 respectively. The spin-polarization is approximately 58 % for $x$=0.03 Mn doping. Dependence of intrinsic spin-polarization on the concentration of magnetic doping (Mn) is shown in Figure 5(b). The transport spin-polarization is reduced to 51.9 % with increase of Mn doping to $x$=0.05. Further increase in doping results in suppression of transport spin-polarization to an approximate value of 48%.

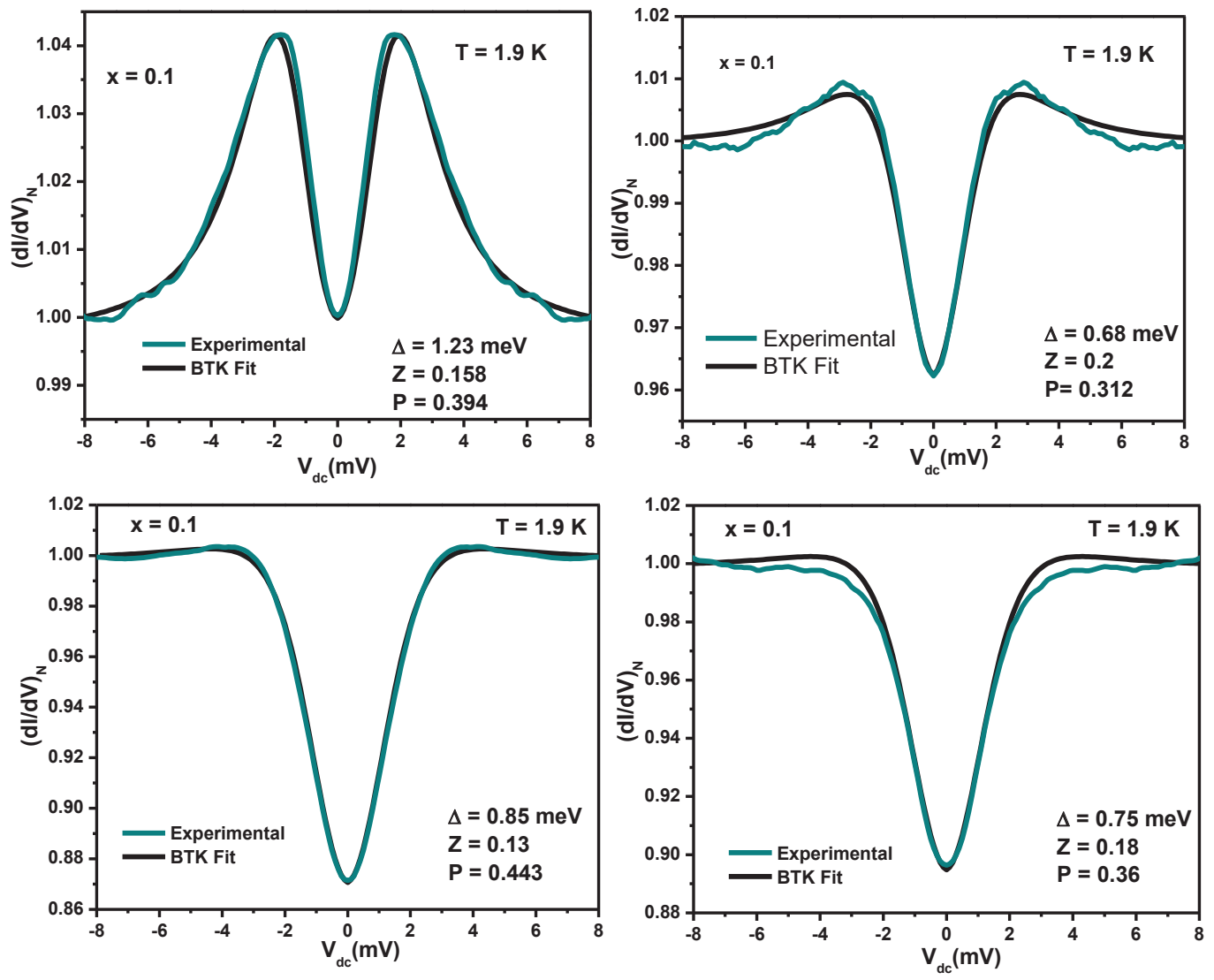

Figure 4: Normalized dI/dV spectra for point-contacts on $Mn_xBi_{2-x}Se_3$ with $x = 0.1$ using a Nb tip. The black lines show BTK fits with spin-polarization included.

It is interesting to note that the value of spin-polarization consistently decreases with the increase in Mn concentration in $Mn_xBi_{2-x}Se_3$. It is rather surpris-ing that while a ferromagnetic phase emerges with Mn doping, the transport spin-polarization decreases. This apparently non-intuitive behavior can be understood by considering the simultaneous destruction of topologically protected surface states with incorporation of magnetic moments in the doped systems. As Mn concentration is increased the topological nature of the surface states get systematically suppressed due to increasing breakdown of time reversal symmetry. This effect also causes a reduction of spin-polarization of the topological fraction of the surface states. On the other hand the emergence of the ferromagnetic phase is accompanied by the magnetization driven spin-polarization. The resultant spin-polarization that is measured in our experiments is a net effect of these two competing processes.

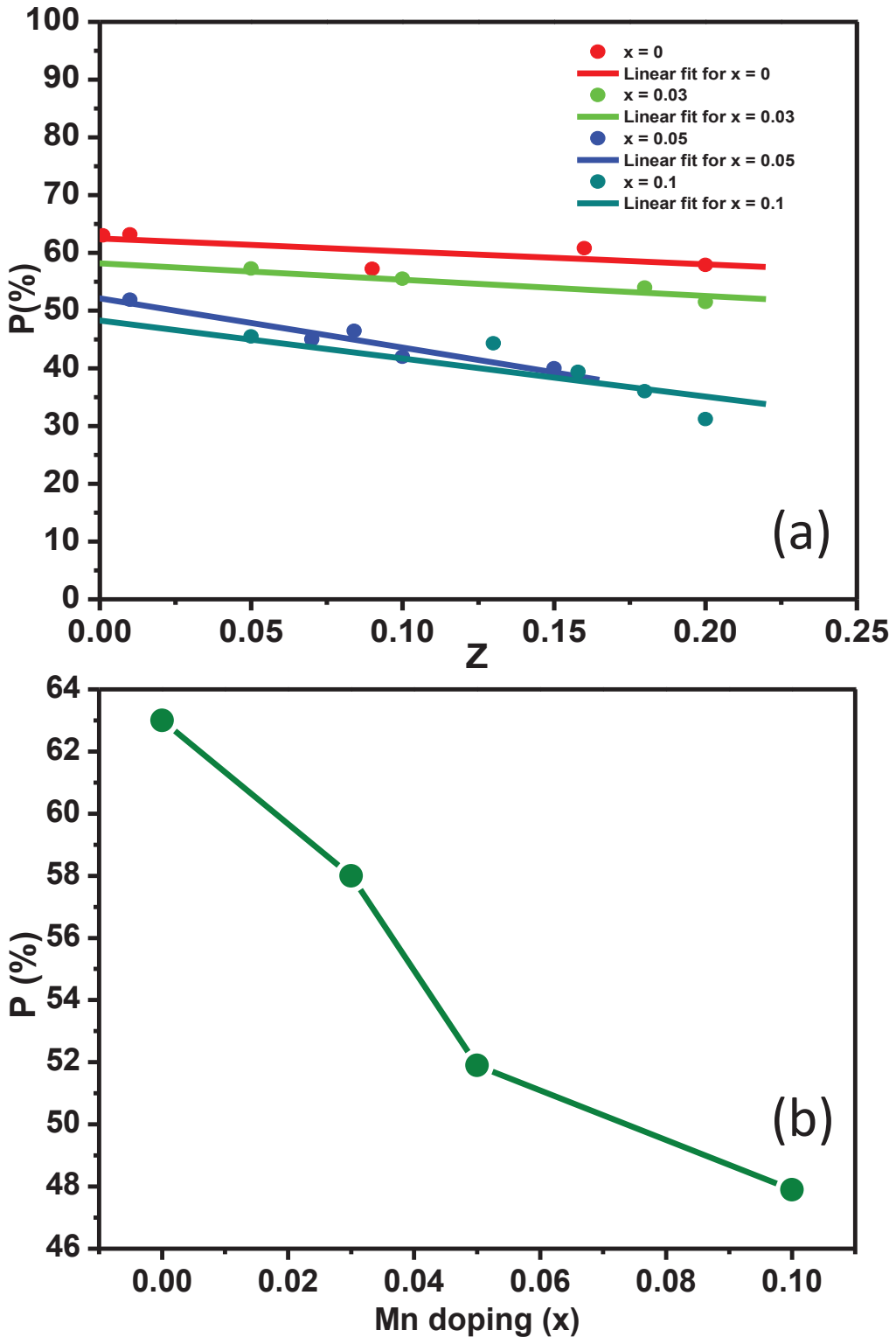

Figure 5: (a) Spin-polarization (P) vs. barrier strength (Z) plot. The solid lines show extrapolation to Z = 0 where the spin-polarization approaches 63 % for the parent sample $Bi_2Se_3$ and gets suppressed in the doped samples $Mn_xBi_{2-x}Se_3$ ($x$= 0.03, 0.05, 0.1). (b) Change in intrinsic spin-polarization with change in Mn concentration ($x$), where $x$ = 0, 0.03, 0.05, 0.1. The lines are guide to the eye.

In conclusion, we have presented scanning tunneling spectroscopy and spin-polarized PCAR spectroscopic studies on single crystals of $Mn_xBi_{2-x}Se_3$ (x= 0, 0.03, 0.05, 0.1) in order to probe the evolution of spin dependent transport properties of $Bi_2Se_3$ with incorporation of magnetic doping. $Bi_2Se_3$ exhibits a high transport spin-polarization of approximately 63 %. Doping with Mn results in a suppression of transport spin-polarization down to 48% with 10% Mn doping. Differential conductance (dI/dV *vs.* V) spectra in $Bi_2Se_3$ from STS measurements reveal a Dirac cone with a Dirac point around 100 mV below the Fermi energy, whereas Mn doping in $Bi_2Se_3$ resulted in disappearance of the Dirac cone and opening of a semiconducting gap in the STS spectra. The reduction in transport spin-polarization can be attributed to the breakdown of time reversal symmetry in the doped system with emergence of ferromagnetism. The resulting spin-polarization of 48% in 10% Mn doped $Bi_2Se_3$ is due to the spin-polarization of the ferromagnetic phase after complete disappearance of the topological protection of the surface states. Extension of the current investigations using other magnetic dopants in this and various other families of topological insulators might lead to better understanding of the magnetically doped TIs for unlocking further novel quantum phenomena to pave the way for future applications for example, in spintronics.

We thank Tanmoy Das, Indian Institute of Science, Bangalore, India for fruitful discussions. GS acknowledges financial support from the research grants of (a) Swarnajayanti fellowship awarded by the Department of Science and Technology (DST), Govt. of India under the grant number DST/SJF/PSA-01/2015-16 and (b)from SERB under the grant number EMR/2015/001650.

* Electronic address: `roychowdhury.rajeswari@gmail.com`
† Electronic address: `goutam@iisermohali.ac.in`

)